\documentclass[aps,pra,twocolumn,letterpaper]{revtex4}

\usepackage{amsfonts, amssymb, amsmath, graphicx, bbold}

\DeclareFontFamily{OT1}{pzc}{}
\DeclareFontShape{OT1}{pzc}{m}{it}{<-> s * [1.30] pzcmi7t}{}
\DeclareMathAlphabet{\mathpzc}{OT1}{pzc}{m}{it}

\newcommand{\vect}[1]{\boldsymbol{\mathrm{#1}}}
\newcommand{\ket}[1]{|#1\rangle}

\newcommand{\expval}[1]{\langle#1\rangle}
\newcommand{\abs}[1]{\left\lvert#1\right\rvert}
\newcommand{\Eqref}[1]{Eq.~\eqref{#1}}

\def\mathi{\textrm{i}}

\def\beq{\begin{equation}}
\def\eeq{\end{equation}}
\def\bes{\begin{equation*}}
\def\ees{\end{equation*}}
\def\bfig{\begin{figure}}
\def\efig{\end{figure}}

\def\ie{i.e., }

\def\eg{e.g., }

\def\prl#1#2#3{Phys.\ Rev.\ Lett.\ {\bf #1}, #2 (#3)}
\def\pra#1#2#3{Phys.\ Rev.\ A {\bf #1}, #2 (#3)}
\def\prb#1#2#3{Phys.\ Rev.\ B {\bf #1}, #2 (#3)}
\def\rmp#1#2#3{Rev.\ Mod.\ Phys.\ {\bf #1}, #2 (#3)}

\begin{document}
	\title{Correlations in lowest Landau level vortex states}
	\author{Soheil Baharian}
	\author{Gordon Baym}
	\affiliation{Department of Physics, University of Illinois at Urbana-Champaign, 1110 W. Green St., Urbana, IL 61801, USA}
	\date{\today}
	
	\begin{abstract}
		We show how the configuration-space form of the Bogoliubov ground state wave function of a bosonic condensate with a single vortex in a harmonic trap can be described in terms of bosonic Jastrow correlations. We then generalize this result to study the first effects of such correlations on a mean-field vortex lattice state and show that the included correlations lower the energy below that of the mean-field state. Although the reduction is relatively small, it is a precursor of the more general expected effect of correlations in describing the melting of the vortex lattice at large angular momentum per particle.
	\end{abstract}

	\maketitle

	\section{Introduction}
		With increasing rotation, the ground state of an ultracold gas of bosons in a harmonic trap undergoes a transition from a vortex lattice with broken rotational symmetry, for which mean-field theory provides a good description (see Ref.~\cite{Cooper-ReviewArticle} and references therein), to a series of symmetry-restored and strongly correlated states~\cite{Cooper-PRL-Melting}, bosonic analogs of quantum Hall states~\cite{CorrelatedStates}. This transition is mediated by the correlations present in the interacting system which are absent in mean-field theory.  Towards understanding how these correlations lead to more favorable states, we studied in Ref.~\cite{BaharianBaym-VortexFluctuations} a condensate with a single vortex; including Bogoliubov fluctuations around the mean-field ground state, we showed that the correlations induced by these fluctuations lower the energy of the Bogoliubov ground state compared to that of the mean-field ground state and cause an uncertainty in the position of the vortex.

		Here we take a first step in generalizing this earlier result to a vortex lattice, showing how correlating two particles in the wave function lowers the energy of the lattice. We focus on correlations described by simple bosonic Jastrow factors, $(z_i-z_j)^2$, in the wave function, where $z \sim x + \mathi y$ is the position of a particle in the complex plane. Such correlations tend to lower the interaction energy by keeping the particles apart and are, hence, favored by repulsive interactions. On the other hand, each factor carries two units of angular momentum and, therefore, tends to increase the kinetic energy of the system. With increasing angular momentum, more and more of these factors enter the wave function, and the states become more strongly correlated, \eg as in the Read-Rezayi~\cite{Read-Rezayi} and the bosonic Laughlin~\cite{WilkinGunnSmith-AttractiveBosons} states. The evolution of the system as its angular momentum increases towards and beyond the melting transition and the role that Jastrow correlations play in this phase transition is still an open problem~\cite{Cooper-ReviewArticle}. With increasing angular momentum, particles begin to occupy single-particle states which previously were empty (or had vanishingly small occupations); this increase in the size of the configuration space of the particles can lead to possible quasi-degeneracies between states with different single-particle occupations and, consequently, to the onset of quantum fluctuations which ultimately destroy the vortex lattice. Even for low angular momenta, where only a few vortices are present, correlations still play a significant role in redistributing the particles among single-particle states, as Cremon \emph{et al.} find~\cite{Cremon} by studying few-vortex ($N_{\textrm{v}} \le 4$) systems numerically and comparing the exact and mean-field ground states. This paper sheds a complementary light on such redistributions.

		An informative example of how correlations function is the gas of attractive bosons studied in Ref.~\cite{WilkinGunnSmith-AttractiveBosons}. For a total (arbitrary) angular momentum $\hbar L$, the ground state wave function is
		\beq
			\psi_1(\vect{z}; L) = z^L_c
			\label{attractiveBosons-GS}
		\eeq
		where $z_c = \sum^N_{i=1}z_i/N$ is the center of mass coordinate. This state has a total interaction energy $\sim -N(N-1)/2$. Moreover, the wave function
		\beq
			\psi_2(\vect{z}; L) = \sum_{i_1 < i_2}(z_{i_1}-z_{i_2})^2 \, \psi_1(\vect{z}; L-2)
			\label{attractiveBosons-ES}
		\eeq
		describes an excited state of the system with the same angular momentum and with a higher interaction energy $\sim -N(N-2)/2$. If we now change the nature of the interactions from attractive to repulsive, these two states switch places in the energy spectrum, with $\ket{\psi_2(L)}$ becoming lower in energy than $\ket{\psi_1(L)}$, although it does not become the ground state. Similar to the Bogoliubov single-vortex state, the Jastrow correlations included in $\ket{\psi_2(L)}$ help to lower the now repulsive interaction energy.

		In this paper, we first show how the real-space form of the Bogoliubov ground state of the single-vortex condensate~\cite{BaharianBaym-VortexFluctuations} includes two-particle bosonic Jastrow factors and can be expanded as a sum over symmetric polynomials with successive number of Jastrow factors. The first term of the sum is just the original uncorrelated mean-field wave function, and the last term has $N/2$ simultaneous Jastrow factors. The effect of such Jastrow correlations is to reduce the total energy by a term $\mathcal{O}(N^{-1})$ which, although small, is a precursor of the more general expected effect of correlations. We then generalize the correlated single-vortex case to a vortex lattice, initially described as a mean-field condensate. Again, we find that the included Jastrow correlations lead to a relative reduction of the energy $\mathcal{O}(N^{-1})$. We also find that the inclusion of these correlations in the trial wave function leads to a nonvanishing density at the vortex cores, indicating the presence of quantum fluctuations of the vortices, similar to the case of the single-vortex system we previously studied~\cite{BaharianBaym-VortexFluctuations}.

		In the next section, we delineate the basic model describing a condensate in terms of Landau levels. In Sec.~\ref{correlationForm}, we expand the Bogoliubov ground state in terms of a series of $N$-particle Fock states with increasing number of particles in the two single-particle states connected to the mean-field ground state through the interactions; we show that these Fock states are represented by \emph{monomial symmetric polynomials} (which, in turn, can be expanded in terms of other symmetric polynomials containing Jastrow factors) and find the form of the correlations present in the wave function. In Sec.~\ref{discuss}, we generalize this construction to a vortex lattice system. Finally, in Appendix~\ref{appendixMonomial}, we derive a general algebraic identity connecting the monomial symmetric polynomials encountered in this problem to symmetric polynomials with successive number of Jastrow factors, and in Appendix~\ref{expandTerms}, we lay out the details of the derivations used to arrive at the results of Sec.~\ref{correlationForm}.

	\section{Basic Model}
		\label{model}
		We consider a gas of $N$ bosons of mass $m$ in a harmonic trap of frequencies $\omega_\perp$ in the $x$--$y$ plane and $\omega_z$ in the $z$ direction, rotating around the $z$ axis with angular velocity $\Omega$. We assume weak two-body repulsive interactions of strength $g=4\pi\hbar^2a/m$, where $a$ is the $s$-wave scattering length. The Hamiltonian in the rotating frame is thus
		\begin{align}
			\mathcal{H}^\prime = &\sum^{N}_{i=1} \bigg[\frac{\vect{p}^2_i}{2m} + \frac{1}{2}m \big(\omega_\perp^2 x^2 + \omega_\perp^2 y^2 + \omega_z^2 z^2\big) - \Omega\ell_i\bigg] \notag \\
				&+ g\sum_{i<j}\delta(\vect{r}_i - \vect{r}_j)
		\end{align}
		where $\ell = \hat{\vect{z}}\cdot(\vect{r}\times\vect{p})$ is the angular momentum along the $z$ direction.

		In the limit of fast rotation ($\Omega \lesssim \omega_\perp$) at zero temperature, the gas becomes quasi-two-dimensional and resides in the axial ground state of the harmonic trap. The single-particle eigenstates of the non-interacting system are the Landau levels, $\ket{nm}$, where $n$ is the radial quantum number and $m \ge -n$ is the angular momentum along the rotation axis. The characteristic interaction energy scale is $V_0 = g / [(2\pi)^{3/2} d_\perp^2 d_z]$ where $d_{\perp, z} = \sqrt{\hbar/m\omega_{\perp, z}}$ are the characteristic oscillator lengths in the transverse and axial directions. We assume the interactions to be sufficiently weak that $V_0 \ll 2\hbar\omega_\perp$; therefore, as $\Omega\to\omega_\perp$, the system resides in the lowest-energy ($n=0$) manifold of Landau levels. The wave function of a particle in the lowest Landau level (LLL) with $m$ units of angular momentum, corresponding to the single-particle state $\ket{0m}$, is
		\beq
			\phi_m(z) = \langle z \ket{0m} = \frac{1}{d_\perp\sqrt{\pi m!}} \, z^m e^{-\abs{z}^2 / 2}
		\eeq
		where $z = (x + \mathi y) / d_\perp$ is the dimensionless position in the complex plane. For brevity, we suppress, throughout this paper, the factor $\exp\big[-\sum^N_{i=1}\abs{z_i}^2\!/2\big] / (d_\perp\sqrt{\pi})^N$ common to all $N$-particle LLL wave functions.

	\section{Correlations in the single-vortex Bogoliubov ground state}
		\label{correlationForm}
		We studied in Ref.~\cite{BaharianBaym-VortexFluctuations} the properties of a single-vortex system in the LLL by including small-amplitude Bogoliubov fluctuations about a mean-field condensate in $\ket{01}$. These fluctuations lower the energy of the Bogoliubov ground state by $-NV_0/4$ compared to the mean-field ground state; the relative reduction in the interaction energy is $\mathcal{O}(N^{-1})$. The vortex, which becomes energetically stable~\cite{LinnFetter-PRA, Fetter-RMP} at the critical rotation frequency $\Omega_c = \omega_\perp - NV_0/4\hbar$, is on average slightly off-center by $\mathcal{O}(1/\sqrt{N})$ (in units of $d_\perp$) due to these quantum fluctuations. In this section, we investigate the nature, in real space, of correlations induced by Bogoliubov fluctuations.

		The Bogoliubov ground state of a single-vortex LLL system at $\Omega=\Omega_c$ is~\cite{BaharianBaym-VortexFluctuations}
		\beq
			\ket{\textrm{G}} = \frac{1}{\sqrt{2}} \, e^{- a^\dagger_2 a^\dagger_0/\sqrt{2}}\ket{N_1},
		\eeq
		where $a_m$ annihilates a particle with angular momentum $m$ from the state $\ket{0m}$, and $\ket{N_1}$ is a coherent state with $N_1$ particles condensed in $\ket{01}$, satisfying the eigenvalue equation $a_1\ket{N_1}=\sqrt{N_1}\ket{N_1}$. This wave function does not conserve the particle number. In order to find its form in configuration space, we restrict the number of particles to $N$ (assumed to be even) and project $\ket{\textrm{G}}$ onto the $N$-particle Fock space. This new wave function, $\ket{\textrm{G};N}$, can be approximated as a sum over states with $N-2m$ particles in $\ket{01}$ and $m$ particles in $\ket{00}$ and $\ket{02}$,
		\beq
			\ket{\textrm{G};N} \simeq \frac{1}{\sqrt{2}} \sum^{N/2}_{m=0} \big(-1/\sqrt{2}\big)^m \, \ket{m,N-2m,m},
			\label{G;N}
		\eeq
		where $\ket{n_0, n_1, n_2}$ contains $n_j$ particles in $\ket{0j}$ (with $j=0, 1, 2$). The norm of this wave function is $\langle\textrm{G};N\ket{\textrm{G};N} = 1-2^{-(1+N/2)}$ and approaches unity when $N\to\infty$.

		The first term in the sum ($m=0$) is just the original mean-field many-body ground state, $\langle\vect{z}\ket{0,N,0} \sim z_1 \cdots z_N$ where $\vect{z}=\{z_1, z_2, \dots, z_N\}$. The $m=1$ term includes first-order corrections and yields $\langle\vect{z}\ket{1,N-2,1} \sim \mathcal{P}\big[z_1^0 z_2 \cdots z_{N-1} (z_N^2 / \sqrt{2}) \big]$ where $\mathcal{P}$ denotes the sum of the distinct permutations with respect to the $z_j$'s needed to symmetrize the wave function. After simplifying this expression (details in Appendix~\ref{expandTerms}), we find that the first-order Bogoliubov corrections take \emph{one} pair of particles out of the condensate and correlate them through a bosonic Jastrow factor,
		\beq
			\sum_{i_1<i_2} (z_{i_1}-z_{i_2})^2 \prod_{k \neq i_1, i_2} z_k \equiv J_1(\vect{z}).
			\label{firstJastrowFactor}
		\eeq
		Similarly, the second-order Bogoliubov correction, the $m=2$ term in \Eqref{G;N}, leads to $\langle\vect{z}\ket{2,N-4,2} \sim \mathcal{P}\big[z_1^0 z_2^0 z_3 \cdots z_{N-2} (z_{N-1}^2 / \sqrt{2}) (z_N^2 / \sqrt{2})\big]$. Simplification of the resulting expression (details in Appendix~\ref{expandTerms}) shows that \emph{two} pairs of particles are correlated through two simultaneous Jastrow factors, resulting in the following term in the wave function
		\beq
			\sideset{}{^\prime}\sum \big(z_{i_1} - z_{i_2}\big)^2 \big(z_{i_3} - z_{i_4}\big)^2 \prod_{k \neq i_1 \dots i_4} \mspace{-10mu} z_k \equiv J_2(\vect{z}),
		\eeq
		where the primed sum indicates the constraints $i_1<i_2, \;\; i_3<i_4, \;\; i_1<i_3, \;\; i_2 \neq i_3, i_4$.

		In fact, the real-space projection of the $m^{\textrm{th}}$ term in the expansion \eqref{G;N} has up to $m$ simultaneous Jastrow factors. To see this structure, we recast this term as
		\begin{align}
			\langle\vect{z}\ket{&m,N-2m,m} \notag \\
				&\sim \mathcal{P}\big[z^0_1 \cdots z^0_m (z^2_{m+1}/\sqrt{2})\cdots(z^2_{2m}/\sqrt{2}) z_{2m+1} \cdots z_{N}\big] \notag \\
				&= \frac{1}{2^{m/2}} \Big[\tbinom{N}{N-2m}\tbinom{2m}{m}\Big]^{-\frac{1}{2}} \, \mathpzc{m}_{\{\underbrace{\scriptstyle 0 \dots 0}_{m}, \underbrace{\scriptstyle 2 \dots 2}_{m}, \underbrace{\scriptstyle 1 \dots 1}_{N-2m}\}}(\vect{z}),
			\label{mExpansion}
		\end{align}
		where $\binom{N}{N-2m}\binom{2m}{m}$ is the number of distinct terms produced by the permutations. The monomial symmetric polynomial~\cite{SymFunc} $\mathpzc{m}_{\vect{\alpha}}(\vect{z})$ is defined in Appendix~\ref{appendixMonomial}, and its representation in terms of symmetric polynomials with successive number of Jastrow factors, determined in Appendix~\ref{expandTerms}, is given by \Eqref{m-monomialJastrow}. Thus,
		\begin{align}
			\langle\vect{z}&\ket{m,N-2m,m} = \frac{1}{2^{m/2}} \Bigg[\frac{1}{m!} \sqrt{\frac{N!}{(N-2m)!}} \, J_0(\vect{z}) \notag \\
				&+ \sqrt{\frac{(N-2m)!}{N!}} \, \sum^{m}_{j=1} 2^{j-1} \, \frac{(2m-2j)!}{(m-j)! \, j!} \, J_j(\vect{z})\Bigg]
			\label{m}
		\end{align}
		where the $N$-variable symmetric polynomial $J_j(\vect{z})$, given by \Eqref{J-term}, includes $j$ successive Jastrow factors. We immediately see up to $m$ pairs of Jastrow-correlated particles in the $m^{\textrm{th}}$-order Bogoliubov correction to the mean-field ground state.

		Substituting \Eqref{m} into \Eqref{G;N} and changing the order of summation using the identity $\sum^{N/2}_{m=0} \sum^{m}_{j=0} = \sum^{N/2}_{j=0} \sum^{N/2}_{m=j}$, we finally arrive at the expansion of the Bogoliubov ground state in terms of Jastrow polynomials,
		\beq
			\langle\vect{z}\ket{\textrm{G};N} = \frac{1}{\sqrt{2}} \sum^{N/2}_{j=0} A_j \, J_j(\vect{z}),
			\label{GwithJastrow}
		\eeq
		where
		\begin{align}
			A_0 &= \sum^{N/2}_{m=0} \frac{(-1)^m}{2^m \, m!} \, \sqrt{\frac{N!}{(N-2m)!}} \, , \notag \\
			A_{j \neq 0} &= \sum^{N/2}_{m=j} \frac{(-1)^m \, (2m-2j)!}{2^{m-j+1} \, (m-j)! \, j!} \, \sqrt{\frac{(N-2m)!}{N!}} \, . \notag
		\end{align}
		Equation~\eqref{GwithJastrow} shows how incorporating Bogoliubov fluctuations in the mean-field ground state leads to pairs of particles being forced out of the condensate and correlated in the Jastrow form. The last term in the expansion above has correlations represented by $N/2$ Jastrow factors, and its coefficient is $\mathcal{O}(N^{-N})$ for large $N$.

		Note that in the thermodynamic limit ($N\to\infty$), the mean-field ground state as described by the Gross-Pitaevskii equation is the true ground state of the system (see, \eg Ref.~\cite{Cooper-ReviewArticle}). In fact, the relative reduction in the energy between the Bogoliubov and the mean-field ground states is $\mathcal{O}(N^{-1})$ for large $N$~\cite{BaharianBaym-VortexFluctuations}. In mesoscopic Bose-condensed systems, the role played by the correlations can be significant, with the Bogoliubov wave function energetically favored over the mean-field solution. Moreover, in \Eqref{GwithJastrow}, the ratio of coefficients of successive terms decreases with increasing $j$; most of the reduction in the interaction energy is due to the first term, with only a single Jastrow factor [as in \Eqref{firstJastrowFactor}].

	\section{Extension to mean-field vortex lattices}
		\label{discuss}
		As discussed above, the Bogoliubov ground state $\ket{\textrm{G}}$, through the quantum fluctuations, has a lower energy than the mean-field ground state. It is clear from the form of the Jastrow polynomial $J_j(\vect{z})$ in \Eqref{J-term} that this lower-energy state $\ket{\textrm{G}; N}$ is constructed by correlating $j$ pairs of particles through $j$ distinct Jastrow factors, thereby leaving only $N-2j$ particles in the original mean-field condensate, $\ket{01}$. As a second example of the effect of Jastrow correlations, we argued, using the wave functions studied in Ref.~\cite{WilkinGunnSmith-AttractiveBosons} for a gas of attractive bosons, that correlating two particles through a Jastrow factor, \Eqref{attractiveBosons-ES}, reduces the energy for repulsive bosons.

		We now show that such Jastrow correlations also lower the energy of a vortex lattice state. In mean-field theory, an $N$-particle LLL condensate with $N_{\textrm{v}}$ vortices at the positions $\{\xi_j\}$ (on a triangular lattice) takes the form
		\beq
			\psi_{\textrm{mf}}(\vect{z}; N_{\textrm{v}}) = \prod^N_{i=1} \, \prod^{N_{\textrm{v}}}_{j=1} (z_i - \xi_j).
			\label{vortex}
		\eeq
		For large $N_{\textrm{v}}$, the system is well described by the Thomas-Fermi approximation~\cite{BaymPethick-TF, CooperKomineasRead-TFvsGaussian}, with the Thomas-Fermi radius $R$ and rotation rate $\Omega$ given by the solution of the two equations
		\beq
			(R/d_\perp)^2 = \sqrt{\frac{4 b N V_0}{\hbar(\omega_\perp - \Omega)}} = (\Omega/\omega_\perp) N_{\textrm{v}},
		\eeq
		where $b \simeq 1.158$ is the Abrikosov lattice parameter. The state \eqref{vortex} is not an eigenstate of the total angular momentum operator $\hat{L}$, but has $\expval{\hat{L}} = \hbar N \big[\tfrac{1}{3} (R/d_\perp)^2 - 1\big]$~\cite{CooperKomineasRead-TFvsGaussian}.

		To study the effect of Jastrow correlations on the energetics of the vortex lattice, we construct a trial wave function by removing two particles from the mean-field condensate and simutaneously correlating them, arriving at the  wave function
		\beq
			\psi_{\textrm{tr}}(\vect{z}; N_{\textrm{v}}) = \sum_{i_1 < i_2}(z_{i_1}-z_{i_2})^2 \, \psi_{\textrm{mf}}\big(\vect{z}-\{z_{i_1}, z_{i_2}\}; N_{\textrm{v}}\big)
			\label{trial}
		\eeq
		where $\psi_{\textrm{mf}}\big(\vect{z}-\{z_{i_1}, z_{i_2}\}; N_{\textrm{v}}\big)$ is an $(N-2)$-particle coherent state (with particles $i_1$ and $i_2$ removed) supporting the same vortices as the original state~\eqref{vortex}. Since the Jastrow factors in \Eqref{trial} force the particles away from each other, we expect the cloud for the correlated state to extend further in space compared to the mean-field one; in fact, the correlated state carrying the same total angular momentum as the mean-field one has a radius given by
		\beq
			R_{\textrm{tr}}^2 \simeq R^2 \, (1+4/N).
		\eeq

		The total interaction energy (found after a tedious calculation, details of which are beyond the scope of this paper~\cite{Baharian-prelim}) is
		\beq
			v_{\textrm{tr}} \simeq V_0 (4b\nu/3) (N-8),
		\eeq
		where $\nu=N/N_{\textrm{v}}$ is the filling factor; this result is valid for large filling factors. Including Jastrow correlations in the trial wave function indeed lowers the energy [albeit by a term $\mathcal{O}(N^{-1})$] compared to mean-field vortex lattice state, for which $v_{\textrm{mf}} \simeq V_0 (4b\nu/3) (N-1)$ at the same value of the total angular momentum. The relative change in the interaction energy is similar to that for a single-vortex system as well as that for attractive bosons of Ref.~\cite{WilkinGunnSmith-AttractiveBosons}. Moreover, due to correlations, the average density at the vortex cores is non-zero for the trial state~\eqref{trial}, similar to the behavior found in Ref.~\cite{BaharianBaym-VortexFluctuations} for a single vortex. In the limit of large number of vortices and for $\nu \gg 1$, we find that the density at the vortex core is~\cite{Baharian-prelim} $n_{\textrm{tr}}(\xi_j) \sim \nu^{-1} \abs{\xi_j}^2 e^{-\abs{\xi_j}^2}$ (except for the central vortex).

		We note that a relative $\mathcal{O}(N^{-1})$ change in the energy is not enough, in the thermodynamic limit, to drive the system towards the strongly correlated regime where the vortex lattice melts~\cite{Cooper-PRL-Melting}. A detailed description of the melting of the lattice will involve states with large numbers of Jastrow-like correlations, \eg as in Read-Rezayi states. Therefore, vortex lattice wave functions of the form \eqref{trial} are only good for large filling factors where the Gross-Pitaevskii equation is an excellent approximation.

	\section{Conclusion}
		This work is an initial study of the role of correlations in the ground state of a vortex lattice state, in the regime where the Gross-Pitaevskii equation is a good first description and quantum fluctuations are small. Although the advantages of including such interparticle correlations are clear -- keeping the particles apart and reducing the interaction energy in the system -- the detailed correlations in the exact ground state of the vortex lattice are not known analytically. Quantum fluctuations, driving the system towards a melting transition to strongly correlated quantum Hall states, become more pronounced as the angular momentum per particle approaches $\mathcal{O}(N)$ and the particle density becomes small, underlining the importance of interaction-induced correlations in this transition. The real-space form of the Bogoliubov ground state of a single-vortex condensate in the LLL studied here shows explicitly the Jastrow-like correlations of pairs of particles in this state. The Bogoliubov wave function is a superposition of the original uncorrelated mean-field ground state and correlated states with successive number of Jastrow pairs. As we showed, 	including Jastrow-correlated pairs (similar to those in the single-vortex Bogoliubov wave function) in a LLL system with $N_{\textrm{v}}$ vortices on a triangular lattice lowers the energy compared to the mean-field wave function with no correlations; this state also exhibits non-zero density at the vortex cores, reflecting the quantum uncertainty in the vortex positions. Generally, interparticle interactions lead to the occupation of single-particle states that were originally unoccupied in the mean-field picture, allowing the system to explore larger regions of phase space, as effectively takes place in our trial wave function~\eqref{trial}, as well as in Ref.~\cite{BaharianBaym-VortexFluctuations} in the single-vortex Bogoliubov wave function. The next step needed is a systematic study of the evolution of the populations of the single-particle states of the vortex lattice with increasing angular momentum.

	\begin{acknowledgements}
		This work was supported in part by NSF Grants No. PHY07-01611 and PHY09-69790. Also, S.B. would like to thank Rinat Kedem for a helpful discussion on symmetric polynomials and Akbar Jaefari for his help in preparing the material in Appendix~\ref{appendixMonomial}.
	\end{acknowledgements}

	\appendix
		\begin{widetext}
		\section{Monomials and Jastrow factors}
			\label{appendixMonomial}
			In this Appendix, we define the elementary and monomial symmetric polynomials and find the expansion of the latter polynomials in terms of symmetric polynomials with Jastrow factors. We consider a set of $N$ variables, denoted by $\vect{z} = \{z_1, z_2, \dots, z_N\}$, and a set of $N$ exponents, denoted by $\vect{\alpha} = \{\alpha_1, \alpha_2, \dots, \alpha_N\}$. The elementary symmetric polynomials defined on $\vect{z}$ are
			\bes
				\mathpzc{s}_0(\vect{z}) = 1, \mspace{5mu} \mathpzc{s}_1(\vect{z}) = \sum_{i_1} z_{i_1}, \mspace{5mu} \mathpzc{s}_2(\vect{z}) = \sum_{i_1<i_2} z_{i_1}z_{i_2}, \mspace{5mu} \mathpzc{s}_3(\vect{z}) = \sum_{i_1<i_2<i_3} z_{i_1}z_{i_2}z_{i_3}, \mspace{5mu} \dots, \mspace{5mu} \mathpzc{s}_N(\vect{z}) = \!\! \sum_{i_1 < i_2 < \dots < i_N} \!\! z_{i_1} z_{i_2} \cdots z_{i_N} = \prod_{k} z_k.
			\ees
			The monomial symmetric polynomials, denoted by $\mathpzc{m}_{\vect{\alpha}}(\vect{z})$, are defined as the sum over all $z^{\alpha_{i_1}}_{1} z^{\alpha_{i_2}}_{2} \cdots z^{\alpha_{i_N}}_{N}$ where the exponents $\alpha_{i_1}, \alpha_{i_2}, \dots, \alpha_{i_N}$ range over all distinct permutations one can get from $\vect{\alpha}$~\cite{SymFunc}. For example, for $N=3$, we have $\mathpzc{m}_{\{2,0,0\}}(z_1, z_2, z_3) = z^2_1 z^0_2 z^0_3 + z^0_1 z^2_2 z^0_3 + z^0_1 z^0_2 z^2_3 = z^2_1+z^2_2+z^2_3$.

			The identity $z^2_1 + z^2_2 = (z_1 - z_2)^2 + 2 z_1 z_2$ for $N=2$ can be rewritten in terms of the symmetric polynomials defined above as $\mathpzc{m}_{\{2,0\}}(z_1, z_2) = (z_1 - z_2)^2 + 2 \, \mathpzc{s}_2(z_1 ,z_2)$. There exists a similar identity for $N=4$ [see \Eqref{N=4monomial} below] which, in the language of symmetric polynomials, becomes
			\begin{align}
				\mathpzc{m}_{\{2,2,0,0\}}(z_1, z_2, z_3, z_4) = &\frac{1}{2}\Big[(z_1-z_2)^2 (z_3-z_4)^2 + (z_1-z_3)^2 (z_2-z_4)^2 + (z_1-z_4)^2 (z_2-z_3)^2\Big] \notag \\
					&+ \Big[(z_1 - z_2)^2 z_3 z_4 + (z_1 - z_3)^2 z_2 z_4 + (z_1 - z_4)^2 z_2 z_3 \notag \\
					&\mspace{30mu} + (z_2 - z_3)^2 z_1 z_4 + (z_2 - z_4)^2 z_1 z_3 + (z_3 - z_4)^2 z_1 z_2\Big] \notag \\
					&+ 6 \, \mathpzc{s}_4(z_1, z_2, z_3, z_4).
			\end{align}
			We now find a similar identity for general $N$, assuming, without loss of generality, that $N=2n$. Defining
			\begin{align}
				&\mspace{75mu}\mathpzc{m}_{\{\underbrace{\scriptstyle 2 \dots 2}_{n}, \underbrace{\scriptstyle 0 \dots 0}_{n}\}}(\vect{z}) = \mathcal{P}\big[z^2_1 z^2_2 \cdots z^2_n z^0_{n+1} z^0_{n+2} \cdots z^0_{2n-1} z^0_{2n}\big], \\
				&J_i(\vect{z}) = \mathcal{P}\big[\overbrace{(z_1-z_2)^2 (z_3-z_4)^2 \cdots (z_{2i-1}-z_{2i})^2}^{\textrm{$i$ Jastrow pairs}} z_{2i+1} z_{2i+2} \cdots z_{2n-1} z_{2n}\big] \label{J-term},
			\end{align}
			we can write
			\beq
				\mathpzc{m}_{\{2 \dots 2, 0 \dots 0\}}(\vect{z}) = \sum^{n}_{i=0}c_i \, J_i(\vect{z}).
				\label{MonomialJastrowExpansion}
			\eeq
			Note that $J_0(\vect{z}) = \mathpzc{s}_{2n}(\vect{z})$.

			To find the coefficients, we proceed as follows. First, we set $z_j=1$ for all $j$. Therefore, in the expansion~\eqref{MonomialJastrowExpansion}, only the $c_0$-term is non-zero. Since the number of terms in the monomial is $(2n)!/(n!)^2$ and all are equal to $1$ in this case, we find $c_0 = (2n)!/(n!)^2$. Next, we set $z_1=0$ and $z_{j \neq 1}=1$. We find that on the right side of \Eqref{MonomialJastrowExpansion}, only the $c_1$-term survives if $z_1$ is one of the two variables in the Jastrow pair, while on the left side, only terms with $z_1^0$ survive. There are $\binom{2n-1}{1}$ ways on the right to make a Jastrow pair with $z_1$ and one other variable; the remaining variables can be arranged in only one way. On the left, for terms with $z_1^0$, there are $(2n-1)!/[n!(n-1)!]$ ways to get a non-zero value (which is $1$). Therefore $c_1 = (2n-2)!/[n!(n-1)!]$. Generalizing this approach to find $c_k$ (with $k \leq n$), we set $z_1=z_2=\dots=z_k=0$ and the rest of $z_j$'s equal to $1$ and proceed as before to find
			\beq
				c_k = \frac{(2n-2k)!}{n!(n-k)!} = \frac{(N-2k)!}{(N/2)!(N/2-k)!}
			\eeq
			which yields
			\beq
				\mathpzc{m}_{\{2 \dots 2, 0 \dots 0\}}(\vect{z}) = \sum^{N/2}_{i=0}\frac{(N-2i)!}{(N/2)!(N/2-i)!} \, J_i(\vect{z}).
				\label{MonomialJastrowExpansionCoefficients}
			\eeq

			Let us count the number of terms in each $J_j(\vect{z})$. We write the act of the permutation operator $\mathcal{P}$ as
			\beq
				J_j(\vect{z}) = \sideset{}{^\prime}\sum (z_{i_1}-z_{i_2})^2 (z_{i_3}-z_{i_4})^2 \cdots (z_{i_{2j-1}}-z_{i_{2j}})^2 \prod_{k \neq i_1 \dots i_{2j}}z_k
			\eeq
			where the prime on the sum indicates the following conditions
			\beq
				\begin{split}
					i_1<i_2&, \mspace{20mu} i_3<i_4, \mspace{20mu} \dots, \mspace{20mu} i_{2j-1}<i_{2j}, \\
					&i_1<i_3<i_5<\dots<i_{2j-1}, \\
					i_{2l} \neq i&_{2l+1}, \, i_{2l+2}, \, \dots, \, i_{2j} \mspace{10mu} \textrm{for} \mspace{10mu} 1 \le l < j.
				\label{ConditionsForJ}
				\end{split}
			\eeq
			In order to construct the Jastrow factors, we choose the $z$'s in them as follows. We pick two $z$'s for the first Jastrow factor in $\binom{N}{2}$ distinct ways, then the two different $z$'s for the second factor in $\binom{N-2}{2}$ distinct ways, and so on until the last one for which there are $\binom{N-2j+2}{2}$ distinct ways. Therefore, we have $N! / [(N-2j)! \, 2^j]$ distinct ways to pick the $z$'s for the Jastrow factors. Moreover, the Jastrow factors can be permuted in $j!$ distinct ways among themselves while keeping $J_j(\vect{z})$ invariant; however, only one of these permutations satisfies the constraints above. The remaining $z$'s can be arranged in only one way. Thus, each $J_j(\vect{z})$ has
			\beq
				\frac{N!}{(N-2j)! \, j! \, 2^j}
				\label{TermsInJ}
			\eeq
			distinct terms.

		\section{Expansion terms}
			\label{expandTerms}
			In this Appendix, we discuss the method we use to simplify the expansion terms in the Bogoliubov ground state, \Eqref{G;N}, and to bring out the Jastrow factors that include interparticle correlations. The $m=1$ term in \Eqref{G;N} is proportional to $\mathcal{P}\big[z^0_1 z_2 \cdots z_{N-1} z^2_N\big]$ where the permutations yield $\binom{N}{N-2}\binom{2}{1}$ distinct terms, \ie
			\beq
				\langle\vect{z}\ket{1,N-2,1} = \Big[\tbinom{N}{N-2}\tbinom{2}{1}\Big]^{-1/2} \; \mathcal{P}\big[z^0_1 z_2 \cdots z_{N-1} \big(z^2_N / \sqrt{2}\big)\big].
			\eeq
			To proceed, we note that the indices of summation (and multiplication) are, in fact, dummy variables and find
			\beq
				\sum_{i \neq j} z^0_i z^2_j \prod_{k \neq i,j} z_k = \sum_{i \neq j} \tfrac{1}{2}\big(z^0_i z^2_j + z^2_i z^0_j) \prod_{k \neq i,j} z_k = \sum_{i<j}\big[(z_i - z_j)^2 + 2 z_i z_j\big] \prod_{k \neq i,j}z_k.
			\eeq
			We thus write
			\beq
				\mathcal{P}\big[z^0_1 z_2 \cdots z_{N-1} z^2_N\big] = \sum_{i<j}(z_i-z_j)^2\prod_{k \neq i,j} z_k + 2\tbinom{N}{2}\prod_k z_k
				\label{2ExcitedParticles}
			\eeq
			which leads to the expansion of $\langle\vect{z}\ket{1,N-2,1}$ in terms of Jastrow polynomials.

			The $m=2$ term in \Eqref{G;N} is proportional to $\mathcal{P}\big[z^0_1 z^0_2 z_3 \cdots z_{N-2} z^2_{N-1} z^2_N\big]$ where the permutations yield $\binom{N}{N-4}\binom{4}{2}$ distinct terms, \ie
			\beq
				\langle\vect{z}\ket{2,N-4,2} = \Big[\tbinom{N}{N-4}\tbinom{4}{2}\Big]^{-1/2} \; \mathcal{P}\big[z^0_1 z_2 \cdots z_{N-2} \big(z^2_{N-1} / \sqrt{2}\big) \big(z^2_N / \sqrt{2}\big)\big].
			\eeq
			The permutation operator can be expanded as
			\begin{align}
				\sum_{\substack{i_1 \neq i_2 \\ \neq i_3 \neq i_4}} \frac{1}{2}z^0_{i_1}z^0_{i_2}\frac{1}{2}z^2_{i_3}z^2_{i_4} \prod_{k \neq i_1 \dots i_4} z_k = \frac{1}{4} \sum_{\substack{i_1 \neq i_2 \\ \neq i_3 \neq i_4}} \tfrac{1}{6} \big(&z^0_{i_1}z^0_{i_2}z^2_{i_3}z^2_{i_4} + z^0_{i_1}z^2_{i_2}z^0_{i_3}z^2_{i_4} + z^0_{i_1}z^2_{i_2}z^2_{i_3}z^0_{i_4} \notag \\
					+ &z^2_{i_1}z^0_{i_2}z^0_{i_3}z^2_{i_4} + z^2_{i_1}z^0_{i_2}z^2_{i_3}z^0_{i_4} + z^2_{i_1}z^2_{i_2}z^0_{i_3}z^0_{i_4}\big) \prod_{k \neq i_1 \dots i_4} z_k
				\label{dummyvariables}
			\end{align}
			where the unrestricted sum on the left side overcounts each factor of $z^\alpha_{i_1}z^\alpha_{i_2}$ (with $\alpha=0, 2$) by $2$ (\eg $z^0_1z^0_2$ and $z^0_2z^0_1$); as before, the equality originates from the permutations on the dummy variables $i_1, i_2, i_3, i_4$. The terms in parentheses above can be rewritten in a more suitable form with the identity
			\begin{multline}
				z^2_1 z^2_2 + z^2_1 z^2_3 + z^2_1 z^2_4 + z^2_2 z^2_3 + z^2_2 z^2_4 + z^2_3 z^2_4 = \tfrac{1}{2}\big[(z_1-z_2)^2 (z_3-z_4)^2 + (z_1-z_3)^2 (z_2-z_4)^2 + (z_1-z_4)^2 (z_2-z_3)^2\big] \\
					+ \big[(z_1 - z_2)^2 z_3 z_4 + (z_1 - z_3)^2 z_2 z_4 + (z_1 - z_4)^2 z_2 z_3 + (z_2 - z_3)^2 z_1 z_4 + (z_2 - z_4)^2 z_1 z_3 + (z_3 - z_4)^2 z_1 z_2\big] + 6 z_1 z_2 z_3 z_4.
				\label{N=4monomial}
			\end{multline}
			Using this in \Eqref{dummyvariables} leads to (i) $3$ equal contributions from the first term on the right side of \Eqref{N=4monomial}, each of which leads to a factor of $2 \times 2$ for converting the unrestricted sum to $i_1<i_2$ and $i_3<i_4$ and another factor of $2$ for imposing the condition $i_1<i_3$; and (ii) $6$ equal contributions from the second term on the right side of \Eqref{N=4monomial}, each of which leads to one factor of $2$ for converting the sum to $i_1<i_2$ and another factor of $2$ to count interchangablity of $i_3$ and $i_4$. Therefore,
			\beq
				\sum_{\substack{i_1 \neq i_2 \\ \neq i_3 \neq i_4}} \mspace{-5mu} z^0_{i_1}z^0_{i_2}z^2_{i_3}z^2_{i_4} \mspace{-5mu} \prod_{k \neq i_1 \dots i_4} \mspace{-15mu} z_k = 
				2 \sideset{}{^\prime}\sum \! \big(z_{i_1} - z_{i_2}\big)^2\big(z_{i_3} - z_{i_4}\big)^2 \mspace{-5mu} \prod_{k \neq i_1 \dots i_4} \mspace{-15mu} z_k + 
				4 \sum_{i_1<i_2} (z_{i_1} - z_{i_2})^2 \mspace{-5mu} \prod_{k \neq i_1, i_2} \mspace{-15mu} z_k + \tbinom{N}{4}4!\prod_k z_k \notag
			\eeq
			where the prime on the sum indicates the conditions \eqref{ConditionsForJ}. We now have
			\beq
				\mathcal{P}\big[z^0_1 z^0_2 z_3 \cdots z_{N-2} z^2_{N-1} z^2_N\big] = \frac{1}{2} \sideset{}{^\prime}\sum \! \big(z_{i_1} - z_{i_2}\big)^2\big(z_{i_3} - z_{i_4}\big)^2 \mspace{-5mu} \prod_{k \neq i_1 \dots i_4} \mspace{-15mu} z_k + \sum_{i_1<i_2} (z_{i_1} - z_{i_2})^2 \mspace{-5mu} \prod_{k \neq i_1, i_2} \mspace{-15mu} z_k + \tbinom{N}{4}3!\prod_k z_k
				\label{4ExcitedParticles}
			\eeq
			and, in turn, the expansion of $\langle\vect{z}\ket{2,N-4,2}$ in terms of Jastrow polynomials.

			As shown in \Eqref{mExpansion}, the $m^{\textrm{th}}$ term in~\eqref{G;N} is proportional to the monomial $\mathpzc{m}_{\{0 \dots 0, 2 \dots 2, 1 \dots 1\}}(\vect{z})$ which we rewrite as
			\begin{align}
				\mathpzc{m}_{\{0 \dots 0, 2 \dots 2, 1 \dots 1\}}(\vect{z}) &= \sum_{i_1 \neq \dots \neq i_{2m}} \frac{1}{m!} z^0_{i_1} \cdots z^0_{i_m} \frac{1}{m!} z^2_{i_{m+1}} \cdots z^2_{i_{2m}} \prod_{k \neq i_1 \dots i_{2m}} \mspace{-15mu} z_k \notag \\
					&= \sum_{i_1 \neq \dots \neq i_{2m}} \frac{1}{(m!)^2} \, \frac{1}{(2m)!/(m!)^2} \, \mathpzc{m}_{\{2 \dots 2, 0 \dots 0\}}(z_{i_1} \dots z_{i_{2m}}) \prod_{k \neq i_1 \dots i_{2m}} \mspace{-15mu} z_k \notag \\
					&= \frac{1}{(2m)!} \, \sum^{m}_{j=0} \frac{(2m-2j)!}{m! \, (m-j)!} \left[\sum_{i_1 \neq \dots \neq i_{2m}} J_j(z_{i_1} \dots z_{i_{2m}}) \prod_{k \neq i_1 \dots i_{2m}} \mspace{-15mu} z_k\right]
			\end{align}
			where (i) due to $i_1, \dots, i_{2m}$ being dummy variables, we have used a method similar to~\eqref{dummyvariables} to get the second equality and to represent all the terms in the sum by a new monomial acting on a limited set of $z$'s; and (ii) we use \Eqref{MonomialJastrowExpansionCoefficients} in the last equality.

			To proceed further, we need to recast the square bracket above (which includes Jastrow polynomials defined on the subset $\{z_{i_1}, \dots, z_{i_{2m}}\} \subset \vect{z}$) in terms of Jastrow polynomials acting on the set $\vect{z}$. The result is
			\beq
				\sum_{i_1 \neq \dots \neq i_{2m}} J_j(z_{i_1} \dots z_{i_{2m}}) \prod_{k \neq i_1 \dots i_{2m}} \mspace{-15mu} z_k = \bigg[(2m)!\binom{N}{2m} \, \delta_{j0} + \frac{(2m)! \, 2^{j-1}}{j!} \, (1-\delta_{j0})\bigg] J_j(\vect{z})
			\eeq
			where the details of this derivation are as follows. Clearly, the sum over $i_1, \dots, i_{2m}$ leads to an overcounting which we need to determine separately for each $j$. Since $i_1, \dots, i_{2m}$ are dummy variables, each term in $J_j(z_{i_1} \dots z_{i_{2m}})$ produces the same polynomial after being summed over; this brings in an overcounting factor given by~\eqref{TermsInJ}. On the other hand, since $\sum_{i_1 \neq i_2} = 2\sum_{i_1<i_2}$, due to the conditions \eqref{ConditionsForJ}, we are overcounting by a factor of $2$ for each Jastrow factor (of which there are $j$) and by a factor of $2$ for each two adjacent Jastrow factors (of which there are $j-1$), in toto, an overcounting factor of $2^{2j-1}$. Permutations of the remaining $z$'s outside the Jastrow factors in $J_j(z_{i_1} \dots z_{i_{2m}})$ leave it invariant, and this leads to an overcounting factor of $(2m-2j)!$. Therefore, when we transform $J_j(z_{i_1} \dots z_{i_{2m}})$ to $J_j(z_1 \dots z_N)$, we overcount by a factor of $(2m)! \, 2^{j-1} / j!$ for each $j \neq 0$. For the special case of $j=0$, since there are no Jastrow factors present in $J_0(z_{i_1} \dots z_{i_{2m}})$, we instantly end up with $J_0(z_1 \dots z_N)$ but overcounted by a factor of $(2m)!\,\binom{N}{2m}$.

			Hence, we write the monomial $\mathpzc{m}_{\{0 \dots 0, 2 \dots 2, 1 \dots 1\}}(\vect{z})$ in terms of symmetric polynomials with successive number of Jastrow factors as
			\beq
				\mathpzc{m}_{\{0 \dots 0, 2 \dots 2, 1 \dots 1\}}(\vect{z}) = \frac{1}{m!} \, \sum^{m}_{j=0} \frac{(2m-2j)!}{(m-j)!} \bigg[\binom{N}{2m} \, \delta_{j0} + \frac{2^{j-1}}{j!} \, (1-\delta_{j0})\bigg] J_j(\vect{z})
				\label{m-monomialJastrow}
			\eeq
			and, in turn, find the expansion of $\langle\vect{z}\ket{m,N-2m,m}$ in terms of Jastrow polynomials, \Eqref{m}.
		\end{widetext}

\end{document}